\documentclass[12pt,a4paper]{article}
\usepackage{epsf}
\begin{document}
\title{Production and decay of the neutral top-pion in high energy $e^{+}e^{-}$ colliders}
\author{Chongxing Yue$^{(a,b)}$,
Qingjun Xu$^{b}$, Guoli Liu$^{b}$, Jiantao Li$^{b}$\\ {\small a:
CCAST (World Laboratory) P.O. BOX 8730. B.J. 100080 P.R. China}
\\ {\small b:College of Physics and Information Engineering,}\\
\small{Henan Normal University, Xinxiang  453002. P.R.China}
\thanks{This work is supported by the National Natural Science
Foundation of China, the Excellent Youth Foundation of Henan Scientific
Committee; and Foundation of Henan Educational Committee.}
\thanks{E-mail:cxyue@pbulic.xxptt.ha.cn} }
\date{\today}
\maketitle
\begin{abstract}
\hspace{5mm} We study the production and decay of the neutral
top-pion $\pi_{t}^{0}$ predicted by topcolor-assisted
technicolor(TC2) theory. Our results show that, except the
dominant decay modes $b\bar{b}$, $\bar{t}c$ and $gg$, the
$\pi_{t}^{0}$ can also decay into $\gamma\gamma$ and $Z \gamma$
modes. It can be significantly produced at high energy
$e^{+}e^{-}$ collider(LC) experiments via the processes
$e^{+}e^{-}\rightarrow \pi_{t}^{0}\gamma$ and
$e^{+}e^{-}\rightarrow Z\pi_{t}^{0}$. We further calculate the
production cross sections of the processes
$e^{+}e^{-}\rightarrow\gamma\pi_{t}^{0}\rightarrow\gamma\bar{t}c$
and $e^{+}e^{-}\rightarrow Z\pi_{t}^{0}\rightarrow Z\bar{t}c$. We
find that the signatures of the neutral top-pion $\pi_{t}^{0}$ can
be detected via these processes.
\end {abstract}

\vspace{1.0cm} \noindent
 {\bf PACS number(s)}: 12.60Nz, 14.80.Mz

\newpage
\section{Introduction}
    The cause of electroweak symmetry breaking (EWSB) and the
origin of fermion masses are important problems of current
particle physics. The standard model (SM) accommodates fermion and
weak gauge boson masses by including a fundamental weak doublet of
scalar Higgs bosons. However, the SM can not explain dynamics
responsible for the generation of mass. Furthermore, the scalar
sector suffers from the problems of triviality and unnaturalness.
Thus, the SM can only be an effective field theory valid below
some high energy scale. Topcolor assisted technicolor (TC2) theory
\cite{x1} is an attractive scheme in which there is an explicit
dynamical mechanism for breaking electroweak  symmetry and
generating the fermion masses including the heavy top quark mass.
In TC2 theory, there are no elementary scalar fields and unnatural
or excessive fine-tuning of parameters. Thus, TC2 theory is one of
the important promising candidates for the mechanism of EWSB.

    In TC2 theory\cite{x1}, almost all of the top quark mass
arises from the topcolor interactions. To maintain electroweak
symmetry between top and bottom quarks and yet not generate
$m_{b}\approx m_{t}$, the topcolor gauge group is usually taken to
be a strongly coupled $SU(3)\bigotimes U(1)$. The $U(1)$ provides
the difference that causes only top quark to condense. In order
that topcolor interactions be natural, i.e. without introducing
large isospin violation, it is necessary that EWSB is still mainly
generated by technicolor(TC) interactions. In TC2 theory, extended
technicolor(ETC) interactions are still needed to generate the
masses of light quarks and contribute a few GeV to $m_{t}$, i.e.
$\varepsilon m_{t}$ \cite{x2}. This means that the associated
top-pions $\pi^{0}_{t}$, $\pi^{\pm}_{t}$ are not the longitudinal
bosons $W$ and $Z$, but are separately physically observable
objects. Thus, TC2 theory predicts a number of Pseudo Goldstone
bosons(PGBs) including the technipions in the TC sector and the
top-pions in the topcolor sector. These new particles are most
directly related to the dynamical symmetry breaking mechanism.
Thus, studying the production and decay of these new particles at
high energy colliders will be of special interest.

    The production and decay of the technipions have been
extensively studied in Ref.[3,4]. The virtual effects of the
PGB's(technipions and top-pions) on the processes such as
$qq\rightarrow t\bar{t}$, $gg\rightarrow t\bar{t}$,
$\gamma\gamma\rightarrow t \bar{t}$, $t\rightarrow cV$, and
$e^{+}\gamma\rightarrow tb \bar{\nu_{e}}$ have been discussed in
the literatures, where the signatures and observability of these
new particles were investigated in hadronic colliders\cite{x5},
$\gamma \gamma$ colliders\cite{x6} and $e \gamma$
colliders\cite{x7} . In this paper, we will study the production
and decay of the neutral top-pion at high energy $e^{+}e^{-}$
colliders(LC) and discuss the signatures and observability of the
neutral top-pion.

    The neutral top-pion $\pi_t^0$, as an isospin-triplet, can
couple to a pair of gauge bosons through the top quark triangle
loop in an isospin violating way similar to the couplings of QCD
pion $\pi^{0}$ to a pair of gauge bosons. Our results indicate
that the main contributions to the production cross section of the
neutral top-pion $\pi_{t}^{0}$ are expected to come from the
$e^{+}e^{-}\rightarrow \pi_{t}^{0}\gamma$ channel at the LC
experiments. In this channel, we find that several hundred events
of $\pi_{t}^{0}$ can be produced per year by assuming an
intergrated luminosity $L=100fb^{-1}$. Possibly, a smaller
$\pi_{t}^{0}$ production may also be observed in the
$e^{+}e^{-}\rightarrow \pi_{t}^{0}Z$ channel.

    For TC2 theory, the underlying interactions, topcolor
interactions, are nonuniversal and therefore do not possess a GIM
mechanism. This is an essential feature of this kind of models due
to the need to single out the top quark for condensation. This
nonuniversal gauge interactions result in flavor changing neutral
current(FCNC) vertices when one writes the interactions in the
quark mass eigen-basis. Thus, the top-pions have large Yukawa
coupling to the third family fermions and induce the new flavor
changing scalar couplings including the $t-c$ transitions for the
neutral top-pion $\pi_{t}^{0}$. Within the SM, the flavor changing
processes are controlled by the scalar sector and the one-loop
level FCNCs are very small. Thus, we can detect the neutral
top-pion via the processes $e^{+}e^{-}\rightarrow \pi_{t}^{0}
\gamma \rightarrow \bar{t}c \gamma$ or $e^{+}e^{-}\rightarrow
\pi_{t}^{0}Z\rightarrow \bar{t}cZ$. Our results show that the
production cross sections are significantly large and may be
observable in the future LC experiments.

    In the rest of this paper, we will give our results in
detail. The couplings of the neutral top-pion $\pi_{t}^{0}$ to the
ordinary particles are discussed in Sec.2, and the decay widths of
the relative decay modes are also estimated. In Sec.3 , we
calculate the production cross sections of the processes
$e^{+}e^{-}\rightarrow \pi_{t}^{0}\gamma\rightarrow
\bar{t}c\gamma$ and $e^{+}e^{-}\rightarrow \pi_{t}^{0}Z\rightarrow
\bar{t}cZ$. Our conclusions are given in Sec.4.

\section{The couplings of the neutral top-pion $\pi_{t}^{0}$ to
the ordinary particles}

    To solve the phenomenological difficulties of TC theory, TC2
theory\cite{x1} was proposed by combining TC interactions with the
topcolor interactions for the third generation at the energy scale
of about 1 TeV. In TC2 theory, the TC interactions play a main
role in breaking the electroweak gauge symmetry. The ETC
interactions give rise to the masses of the ordinary fermions
including a very small portion of the top quark mass, namely $
\epsilon m_{t}$ with a model dependent parameter $\epsilon\ll1$.
The topcolor interactions also make small contributions to the
EWSB, and give rise to the main part of the top quark mass,
$(1-\epsilon)m_{t}$, similar to the constituent masses of the
light quarks in QCD. So, for TC2 theory, there is the following
relation:
 \begin{equation}
 \nu_{\pi}^{2}+F_{t}^{2}=\nu_{w}^{2},
 \end{equation}
where $\nu_{\pi}$ represents the contributions of the TC
interactions to the EWSB, $F_{t}=50GeV$ is the top-pion decay
constant, and $\nu_{w}=\nu/\sqrt{2}=174GeV$.

    For TC2 theory, it generally predicts three top-pions with
large Yukawa couplings to the third generation. This induces the
new flavor changing scalar couplings. The couplings of the neutral
top-pion $\pi_{t}^{0}$ to the ordinary fermions can be written
as\cite{x1,x8}:
\begin{equation}
\frac{m_{t}}{\sqrt{2}F_{t}}\frac{\sqrt{\nu_{w}^{2}-F_{t}^{2}}}{\nu_{w}}
[K_{UR}^{tt}K_{UL}^{tt*}\bar{t_{L}}t_{R}\pi_{t}^{0}+K_{UR}^{tc}K_{UL}^{tt*}\bar{t_{L}}c_{R}\pi_{t}^{0}+h.c.],
\end{equation}
where the factor $\frac{\sqrt{\nu_{w}^{2}-F_{t}^{2}}}{\nu_{w}}$
reflects the effect of the mixing between $\pi_{t}^{0}$ and the
would be Goldstone boson \cite{x9}. $K_{UL}^{ij}$ is the matrix
element of the unitary matrix $K_{UL}$ which the CKM matrix can be
derived as $V=K_{UL}^{-1}K_{DL}$ and $K_{UR}^{ij}$ is the matrix
element of the right-handed relation matrix $K_{UR}$. Ref.[8] has
shown that their values can be taken as:
\begin{equation}
K_{UL}^{tt}=1, \hspace{5mm}  K_{UR}^{tt}=1-\epsilon, \hspace{5mm}
K_{UR}^{tc}\leq \sqrt{2\epsilon-\epsilon^{2}}.
\end{equation}
If we assume that the part of the top quark mass generated by the
topcolor interactions makes up $99\%$ of $m_{t}$, i.e.
$\epsilon=0.01$, then we have $K_{UR}^{tc}<0.14$. In the following
calculation, we will take $\epsilon=0.01$ and $K_{UR}^{tc}$ as a
free parameter. The coupling of $\pi_{t}^{0}$ to the bottom quark
can be approximately written as:
\begin{equation}
\frac{m_{b}-m_{b}^{'}}{\sqrt{2}F_{t}}\frac{\sqrt{\nu_{w}^{2}-F_{t}^{2}}}{\nu_{w}}\bar{b}\gamma^{5}b\pi_{t}^{0},
\end{equation}
where $m_b^{'}$ is the ETC generated part of the bottom-quark
mass. According to the idea of TC2 theory, the masses of the first
and second generation fermions are also generated by ETC
interactions. We have $\epsilon
m_{t}=\frac{m_{c}}{m_{s}}m_{b}^{'}$ \cite{x10}. If we take
$m_{s}=0.156GeV$, $m_{c}=1.56GeV$, then we have
$m_{b}^{'}=0.1\times \epsilon m_{t}$.

   The couplings of $\pi_{t}^{0}$ to gauge bosons via the top quark
triangle loop are isospin violating similar to the couplings of
QCD pion $\pi^{0}$ to gauge bosons. The general form of the
effective $\pi_{t}^{0}-B_{1}-B_{2}$ coupling can be written as
\cite{x5}:
\begin{equation}
\frac{1}{1+\delta_{B_{1}B_{2}}}\frac{\alpha S_{\pi_t^0
B_1B_2}}{\pi F_t} \pi_t^0
\epsilon_{\mu\nu\alpha\beta}(\partial^{\mu}
B_{1}^{\nu})(\partial^{\alpha}B_{2}^{\beta}),
\end{equation}
where $B_{1}^{\nu}$ and $B_{2}^{\beta}$ represent the field
operators of the gauge bosons. We define $\alpha$ to be the strong
coupling constant $\alpha_s$ if $B_1$ and $B_2$ are QCD gluons and
equal to $\frac{e^{2}}{4\pi}$ if $B_{1}$ and $B_{2}$ are
electroweak gauge bosons. The anomalous factors
$S_{\pi_t^0B_{1}B_{2}}$ are model dependent. They can be
calculated by using the formulas in Ref.[11]. For the neutral
top-pion $\pi_t^0$, we have:
\begin{equation}
S_{\pi_{t}^{0}g_{a}g_{b}}=\frac{1}{\sqrt{2}}\frac{\sqrt{\nu_{w}^{2}-F_{t}^{2}}}{\nu_{w}}
K_{UR}^{tt}J_{R},
\end{equation}
\begin{equation}
S_{\pi_{t}^{0}\gamma\gamma}=\frac{16}{3\sqrt{2}}
\frac{\sqrt{\nu_{w}^{2}-F_{t}^{2}}}{\nu_{w}}K_{UR}^{tt}J_{R},
\end{equation}
\begin{equation}
S_{\pi_{t}^{0}Z\gamma}=\frac{16}{3\sqrt{2}}
\frac{\sqrt{\nu_{w}^{2}-F_{t}^{2}}}{\nu_{w}}K_{UR}^{tt}\tan\theta_{w}L(m_{\pi_{t}}),
\end{equation}
with
\begin{equation}
L(m_{\pi_{t}})=\int_{0}^{1}dx\int_{0}^{1}dy\left[1+(\frac{m_{\pi_{t}}}{m_{t}})^{2}
x(x-1)y^{2}+(\frac{m_{Z}}{m_{t}})^{2}yx(y-1)\right]^{-1},
\end{equation}
\begin{equation}
J(R_{\pi_{t}^{0}})=\frac{1}{R_{\pi_{t}^{0}}^{2}}\int_{0}^{1}\frac{dx}{x(x-1)}
ln\left[1-R_{\pi_{t}^{0}}^{2}x(1-x)\right].
\end{equation}
Where $R_{\pi_{t}^{0}}=\frac{m_{\pi_{t}}}{m_{t}}$, $\theta_{w}$ is
the Weinberg angle. In the above formulas, we have ignored the
coupling of $\pi_{t}^{0}$ to a pair of gauge bosons $Z$ and taken
$S_{\pi_{t}^{0}ZZ}\approx 0$.

    If we assume that the mass of $\pi_{t}^{0}$ is in the range of $200GeV\leq
m_{\pi_{t}^{0}}\leq350GeV$\cite{x9}, the possible decay modes of
$\pi_{t}^{0}$ are $\bar{t}c$, $b\bar{b}$, $gg$, $\gamma\gamma$ and
$Z\gamma$. Then we have:
\begin{equation}
\Gamma_{total}(\pi_{t}^{0})=\Gamma(\pi_{t}^{0}\rightarrow
b\bar{b})+\Gamma(\pi_{t}^{0}\rightarrow
\bar{t}c)+\Gamma(\pi_{t}^{0}\rightarrow
gg)+\Gamma(\pi_{t}^{0}\rightarrow\gamma\gamma)+\Gamma(\pi_{t}^{0}\rightarrow
Z\gamma).
\end{equation}
Using Eq.(2)----Eq.(9), we can obtain:
\begin{equation}
\Gamma(\pi_{t}^{0}\rightarrow b\bar{b})=\frac{3}{16\pi}
\frac{\nu_{w}^{2}-F_{t}^{2}}{\nu_{w}^{2}}\frac{(m_{b}-0.1\epsilon
m_{t})^{2}}{F_{t}^{2}}m_{\pi_{t}}\sqrt{1-\frac{4m_{b}^{2}}{m_{\pi_{t}^{0}}^{2}}},
\end{equation}
\begin{equation}
\Gamma(\pi_{t}^{0}\rightarrow
\bar{t}c)=\frac{3(1-\epsilon)^{2}}{16\pi}\frac{\nu_{w}^{2}-F_{t}^{2}}{\nu_{w}^{2}}
\frac{m_{t}^{2}m_{\pi_{t}}}{F_{t}^{2}}(K_{UR}^{tc})^{2}\sqrt{1-\frac{m_{t}^{2}}{m_{\pi_{t}^{2}}}},
\end{equation}
\begin{equation}
\Gamma(\pi_{t}^{0}\rightarrow
gg)=\frac{\alpha_{s}^{2}(1-\epsilon)^{2}}{64\pi^{3}}\frac{m_{\pi_{t}}^{3}}{F_{t}^{2}}
\frac{\nu_{w}^{2}-F_{t}^{2}}{\nu_{w}^{2}}J^{2}(R_{\pi_{t}}),
\end{equation}
\begin{equation}
\Gamma(\pi_{t}^{0}\rightarrow\gamma\gamma)=\frac{\alpha_{e}^{2}(1-\epsilon)^{2}}{18\pi^{3}}
\frac{\nu_{w}^{2}-F_{t}^{2}}{\nu_{w}^{2}}\frac{m_{\pi_{t}}^{3}}{F_{t}^{2}}J^{2}(R_{\pi_{t}}),
\end{equation}
\begin{equation}
\Gamma(\pi_{t}^{0}\rightarrow
Z\gamma)=\frac{\alpha_{e}^{2}(1-\epsilon)^{2}}{9\pi^{3}}\frac{\nu_{w}^{2}-F_{t}^{2}}{\nu_{w}^{2}}
\frac{m_{\pi_{t}}^{3}}{F_{t}^{2}}\tan^{2}\theta_{w} \left
(1-\frac{m_{t}^{2}}{m_{\pi_{t}}^{2}}\right)^{2}L^{2}(m_{\pi_{t}}).
\end{equation}

    Since the $\bar{c_{L}}t_{R}$ coupling is very small
\cite{x1, x8}, we have assumed $K_{UR}^{tc}\approx K_{U}^{tc}
=\sqrt{|K_{UL}^{tc}|^{2}+|K_{UR}^{tc}|^{2}}$ in Eq.(13). Using
above equations, we can estimate the branching ratios of the
various decay modes of the neutral top-pion $\pi_{t}^{0}$. In
Fig.1, we plot the resulting branching ratios as a function of
$m_{\pi_{t}}$ for $K_{UR}^{tc}=0.05$. The branching ratios of
anomalous channels $\pi_{t}^{0}\rightarrow \gamma\gamma$ and
$\pi_{t}^{0}\rightarrow Z \gamma$ are very small, so we do not
give the values of $B_{r}(\gamma\gamma)$ and $B_{r}(Z\gamma)$ in
Fig.1. From Fig.1, we can see that the largest branching ratio is
that of the decay channel $\pi_{t}^{0}\rightarrow\bar{t}c$. The
$\pi_{t}^{0}\rightarrow\bar{t}c$ branching ratio varies between
$46\%$ and $65\%$ for its mass in the range of $200GeV$-$350GeV$.
For large value of the $\pi_{t}^{0}$ mass, a considerable ratio of
the $\pi_{t}^{0}$ decay mode can also occur in the anomalous
channel $\pi_{t}^{0}\rightarrow gg$. For $m_{\pi_{t}}=300GeV$, we
have $B_{r}(gg)\simeq 16\%$.

    To see the effect of the parameter $K_{UR}^{tc}$ on the branching
ratio $B_{r}(\bar{t}c)$, we plot $B_{r}(\bar{t}c)$ versus
$K_{UR}^{tc}$ in Fig.2 for $m_{\pi_{t}}=300GeV$. From Fig.2 we can
see that $B_{r}(\bar{t}c)$ is sensitive to the value of parameter
$K_{UR}^{tc}$. For $\epsilon>0.01$, we have
$B_{r}(\bar{t}c)\geq99\%$.

\section{the $\pi_{t}^{0}$ production at high energy $e^{+}e^{-}$ colliders}

    From above discussion, we can see that $gg$ is one of the dominant
decay modes of the neutral top-pion $\pi_{t}^{0}$. This will make
$\pi_{t}^{0}$ observed in the $\bar{t}c$ final state at a hadron
collider very easy. Ref.[8] has extensively studied the signatures
of the $\pi_{t}^{0}$ at the hadron colliders. In this section we
will focus on the $\pi_{t}^{0}$ production at the high energy
$e^{+}e^{-}$ colliders(LC). In the LC experiments, the
$\pi_{t}^{0}$ production channels are represented by the processes
\begin{equation}
e^{+}e^{-}\rightarrow\pi_{t}^{0}\gamma, \hspace{5mm}
e^{+}e^{-}\rightarrow\pi_{t}^{0}Z.
\end{equation}

    For the processes in Eq.(17), the general form of the total cross
section has been given in Ref.[12]. For the neutral top-pion
$\pi_{t}^{0}$, we have assumed that its mass is in the range of
$200GeV\leq m_{\pi_{t}}\leq 350GeV$ and $S_{\pi_{t}^{0}ZZ}=0$.
Thus the production cross sections of the processes in Eq.(17) can
be written as:
\begin{eqnarray}
\sigma(e^{+}e^{-}\rightarrow\pi_{t}^{0}\gamma)&=&\frac{\alpha^{3}}{24\pi^{2}F_{t}^{2}}
(1-\frac{m_{\pi_{t}}^{2}}{S})^{3}
\\ \nonumber
&&[S_{\pi_{t}^{0}\gamma\gamma}^{2}
+\frac{(1-4S_{W}^{2}+8S_{W}^{4})}{8S_{W}^{2}C_{W}^{2}}
\frac{S_{\pi_{t}^{0}\gamma
Z}^{2}}{(1-\frac{m_{Z}^{2}}{S})^{2}}+\frac{1-4S_{W}^{2}}{2S_{W}C_{W}}\frac{S_{\pi_{t}^{0}\gamma
\gamma}S_{\pi_{t}^{0}\gamma Z}}{1-\frac{m_{Z}^{2}}{S}}],
\end{eqnarray}
\begin{equation}
\sigma(e^{+}e^{-}\rightarrow\pi_{t}^{0}Z)=\frac{\alpha^{3}}{24\pi^{2}F_{t}^{2}}
(1-\frac{m_{\pi_{t}}^{2}}{S}-\frac{m_{Z}^{2}}{S})^{3}S_{\pi_{t}^{0}Z\gamma}^{2},
\end{equation}
where $\sqrt{S}$ is the center-of-mass energy. Using the above
equations, we can calculate the $\pi_{t}^{0}$ production cross
sections via the processes
$e^{+}e^{-}\rightarrow\gamma\pi_{t}^{0}$ and
$e^{+}e^{-}\rightarrow Z\pi_{t}^{0}$. Our results are shown in
Fig.3 as a function of $m_{\pi_{t}}$ for $\sqrt{S}=500GeV$,
$\epsilon=0.01$. From Fig.3, we can see that the production cross
sections increase with decreasing the value of $m_{\pi_{t}}$ in
most of the parameter space and the cross section
$\sigma_{\gamma\pi_{t}^{0}}$ is larger than the cross section
$\sigma _{Z \pi_{t}^{0}}$. However, the cross section $\sigma_{
\gamma\pi_{t}^{0}}$ increases with increasing the mass
$m_{\pi_{t}}$ for $m_{\pi_{t}}>320GeV$. The reason is that the
factor $S_{\pi_{t}^{0}\gamma\gamma}$ increases speedily as
$R_{\pi_{t}^{0}}=\frac{m_{\pi_{t}}}{m_{t}}$ goes to
$R_{\pi_{t}^{0}}=2$. For $m_{\pi_{t}}=300GeV$, the cross sections
are $\sigma_{\gamma\pi_{t}^{0}}=2.5fb$ and $\sigma_{Z
\pi_{t}^{0}}=0.47fb$. To see the effects of the center-of-mass
energy $\sqrt{S}$ on the production cross sections, we plot the
cross section $\sigma_{\pi_{t}^{0}\gamma(Z)}$ versus $\sqrt{S}$ in
Fig.4 for $\epsilon=0.01$ and $m_{\pi_{t}}=300GeV$. From Fig.4, we
can see that the cross sections increase with increasing the value
of $\sqrt{S}$. For the process
$e^{+}e^{-}\rightarrow\pi_{t}^{0}\gamma$, the cross section
increases from $2.5fb$ to $7.1fb$ as $\sqrt{S}$ increases from
$500GeV$ to $1000GeV$.

    We consider two future $e^{+}e^{-}$ collider scenarios: a LC with
$\sqrt{S}=500GeV$ and a yearly intergrated luminosity of
$L=50fb^{-1}$ and a LC with $\sqrt{S}=1000GeV$ and $L=100fb^{-1}$.
The yearly production events can be easily calculated. Our
numerical results are shown in Table 1. From Table 1 we can see
that the neutral top-pion $\pi_{t}^{0}$ can be significantly
produced via the $e^{+}e^{-}\rightarrow\pi_{t}^{0}\gamma$ channel
at the future high energy $e^{+}e^{-}$ colliders. For this
channel, several hundred events of $\pi_{t}^{0}$ can be produced
per year which would be observed at the LC experiments,
$\pi_{t}^{0}$ can also be produced via the
$e^{+}e^{-}\rightarrow\pi_{t}^{0}Z$ channel. Several ten events of
$\pi_{t}^{0}$ can be produced per year which may be observed at
the LC experiments with $\sqrt{S}=1000GeV$.
\begin{table}[hbt]
\begin{center}
\caption{The number of $\pi_t^0$ produced per year
 from $\pi_t^0 \gamma$ and $\pi_t^0 Z$
 at the LC experiments}
 \label{ai:bd}
  \vspace{0.5cm}
\begin{tabular}{|c|c|c|c|}
\hline

 $\sqrt{s}$   &$m_{\pi_t}=250GeV$   &$m_{\pi_t}=300GeV$
 &$m_{\pi_t}=350GeV$ \\ \cline{2-4}
              &$\left\{\pi_{t}^{0}\gamma, \pi_{t}^{0}Z\right\}$ & $\left\{\pi_{t}^{0}\gamma,
              \pi_{t}^{0}Z\right\}$
              & $\left\{\pi_{t}^{0}\gamma, \pi_{t}^{0}Z\right\}$ \\ \hline
 $500GeV$     &150, 31 &125, 23 &163, 22 \\
  \hline
 $1000GeV$    &582, 136 &713, 153 &1650, 269\\
 \hline
\end{tabular}
\end{center}
\end{table}

    In the SM, $\bar{t}c$ pair production cross sections are
unobservably small due to the unitarity of CKM matrix \cite{x13}.
Thus, any signal of such $t-c$ transitions will be a clear
evidence of new physics beyond the SM. This fact has led to a lot
of theoretical activities involving $t-c$ transitions within some
specific popular models beyond the SM. For example, $\bar{t}c$
production has been studied in Multi Higgs doublets models \cite
{x14}, in supersymmetry with R-Parity violation \cite{x15}, in
models with extra vector-like quarks \cite{x16} and model
independent\cite{x17}. For TC2 models, we have discussed the
contributions of the new gauge bosons and the neutral top-pion
$\pi_{t}^{0}$ to $\bar{t}c$ production\cite{x18}. From our
discussion in Sec.2, we can see that the $\bar{t}c$ mode is one of
the dominant decay modes of the neutral top-pion $\pi_{t}^{0}$.
Thus the new flavor changing
 scalar coupling $\pi_{t}^{0}\bar{t}c$ may have significant
contributions to the processes $e^{+}e^{-}\rightarrow
\gamma\pi_{t}^{0}\rightarrow\gamma\bar{t}c$ and
$e^{+}e^{-}\rightarrow Z \pi_{t}^{0}\rightarrow Z \bar{t}c$. The
neutral top-pion $\pi_{t}^{0}$ may be detected by studying these
processes.
    Since we only consider the real $\pi_{t}^{0}$ production, the
cross sections of the processes
$e^{+}e^{-}\rightarrow\gamma\bar{t}c$ and $e^{+}e^{-}\rightarrow Z
\bar{t}c$ contributed by $\pi_{t}^{0}$ can be written as:
\begin{equation}
\sigma_{\gamma\bar{t}c}\approx\sigma(e^{+}e^{-}\rightarrow\gamma\pi_{t}^{0})\times
B_{r}(\bar{t}c),
\end{equation}
\begin{equation}
\sigma_{Z\bar{t}c}\approx\sigma(e^{+}e^{-}\rightarrow Z
\pi_{t}^{0})\times B_{r}(\bar{t}c),
\end{equation}
with
\begin{equation}
B_{r}(\pi_{t}^{0}\rightarrow\bar{t}c)=\frac{\Gamma(\pi_{t}^{0}\rightarrow
\bar{t}c)}{\Gamma_{total}(\pi_{t}^{0})}.
\end{equation}
Where $\Gamma_{total}(\pi_{t}^{0})$ is the total decay width of
the $\pi_{t}^{0}$. In Fig.5, we plot the cross sections
$\sigma_{\gamma\bar{t}c}$ and $\sigma_{Z\bar{t}c}$ as a function
of the mass $m_{\pi_{t}}$ for $\sqrt{S}=500GeV$, in which the
solid and dotted lines stand for the final states $\gamma\bar{t}c$
and $Z\bar{t}c$, respectively. From Fig.5 we can see that, for
$200GeV\leq m_{\pi_{t}}\leq 350GeV$, the cross section
$\sigma_{\gamma\bar{t}c}$ varies from $1.29fb$ to $2.05fb$ and
$\sigma_{Z\bar{t}c}$ varies from $0.20fb$ to $0.45fb$. Thus, for a
LC with $\sqrt{S}=500GeV$, TC2 models predict tens and up to one
hundred of $\gamma\bar{t}c$ raw events, and only few of
$Z\bar{t}c$ raw events. Detecting the neutral top-pion
$\pi_{t}^{0}$ is very difficult via the process
$e^{+}e^{-}\rightarrow Z \pi_{t}^{0}\rightarrow Z \bar{t}c$.
However, we can study the signature and observability of
$\pi_{t}^{0}$ via the process
$e^{+}e^{-}\rightarrow\gamma\pi_{t}^{0}\rightarrow\gamma\bar{t}c$
and further test TC2 models. In Fig.6 we show the
$\sigma_{\gamma\bar{t}c}$ and $\sigma_{Z\bar{t}c}$ as a function
of $\sqrt{S}$ for $m_{\pi_{t}}=300GeV$. Same as Fig.5, the solid
and dotted lines stand for the final states $\gamma\bar{t}c$ and
$Z\bar{t}c$, respectively. From Fig.6, we can see that the cross
sections increase with increasing $\sqrt{S}$. For a LC with
$\sqrt{S}=1000GeV$ and $L=100fb^{-1}$, there can be about 443
$\gamma\bar{t}c$ raw events and 95 $Z\bar{t}c$ raw events for
$m_{\pi_{t}}=300GeV$. Thus, the neutral top-pion $\pi_{t}^{0}$ can
also be detected via the process $e^{+}e^{-}\rightarrow Z
\bar{t}c$ at a LC with $\sqrt{S}=1000GeV$.

\section{summary}

    To avoid or solve the problems, such as triviality and
unnaturalness arising from the elementary Higgs field, various
kinds of dynamical electroweak symmetry breaking models have been
proposed, and among which TC2 theory is one of the important
candidates. TC2 theory predicts the existence of the top-pions.
These new particles are most directly related to the dynamical
symmetry breaking mechanism and can be seen as the characteristic
feature of TC2 theory. Thus it is of great importance to
understand what about the prospects for discovering the top-pions
and whether we can study their properties to determine key
features and parameters of TC2 theory.

    In this paper, we have studied the production and decay of
the neutral top-pion $\pi_{t}^{0}$ at high-energy $e^{+}e^{-}$
colliders. The top-pions have large Yukawa couplings to the third
generation fermions and induce the new flavor changing scalar
couplings including the $\pi_{t}^{0}\bar{t}c$ coupling for the
neutral top-pion $\pi_{t}^{0}$. The neutral top-pion
$\pi_{t}^{0}$, as an isospin-triplet, can couple to a pair of
gauge bosons through the top quark triangle loop in an isospin
violating way similar to the couplings of QCD pion $\pi^{0}$ to a
pair of gauge bosons. Our results indicate that the dominant decay
modes are $\bar{t}c$ and $gg$. Thus, the $\pi_{t}^{0}$ can be
significantly produced at hadronic colliders. The signatures and
observability of the neutral top-pion $\pi_{t}^{0}$ can be studied
at the Tevatron. This is consistent with the results of Ref.[8].

    Except the decay modes $\bar{t}c$ and $gg$, the $\pi_{t}^{0}$
can also decay into $\gamma\gamma$ and $Z\gamma$ modes. Thus, the
$\pi_{t}^{0}$ can be produced at the LC experiments via the
processes $e^{+}e^{-}\rightarrow\pi_{t}^{0}\gamma$ and
$e^{+}e^{-}\rightarrow Z \pi_{t}^{0}$. For $\sqrt{S}=1000GeV$, the
production cross sections $\sigma_{\pi_{t}^{0}\gamma}$ and
$\sigma_{\pi_{t}^{0}Z}$ can reach $7.1fb$ and $1.5fb$,
respectively. It is expected that there will be several hundreds
$\gamma\pi_{t}^{0}$ and $Z\pi_{t}^{0}$ events per year at the LC
experiments with $\sqrt{S}=1000GeV$. Thus the neutral top-pion
$\pi_{t}^{0}$ will be observed at the high energy $e^{+}e^{-}$
colliders.

    Since the $\bar{t}c$ production cross sections are unobservably
small in the SM, any signal of $\bar{t}c$ production will be a
clear evidence of new physics beyond the SM. The flavor changing
scalar coupling $\pi_{t}^{0}\bar{t}c$ has significant
contributions to the processes
$e^{+}e^{-}\rightarrow\gamma\pi_{t}^{0}\rightarrow\gamma\bar{t}c$
and $e^{+}e^{-}\rightarrow Z \pi_{t}^{0}\rightarrow Z \bar{t}c$.
The neutral top-pion $\pi_{t}^{0}$ can be detected by studying
these processes. Furthermore, we can test TC2 models via these
processes at the high-energy $e^{+}e^{-}$ colliders.
\newpage
\vskip 2.0cm
\begin{center}
{\bf Figure captions}
\end{center}
\begin{description}
\item[Fig.1:]The branching ratios of the neutral top-pion as
a function of its mass $m_{\pi_{t}}$.
\item[Fig.2:]The branching ratio
$B_{r}(\pi_{t}^{0}\rightarrow\bar{t}c)$ versus the free parameter
$K_{UR}^{tc}$ for the mass $m_{\pi_{t}}=300GeV$.
\item[Fig.3:]The production cross sections $\sigma_{\gamma\pi_{t}^{0}}$ and $\sigma_{Z \pi_{t}^{0}}$ of the neutral
top-pion $\pi_{t}^{0}$ as a function of its mass for
$\sqrt{S}=500GeV$.
\item[Fig.4:]The production cross sections $\sigma_{\gamma\pi_{t}^{0}}$ and $\sigma_{Z \pi_{t}^{0}}$ versus the
center-of-mass energy $\sqrt{S}$ for $m_{\pi_{t}}=300GeV$.
\item[Fig.5:]The cross sections $\sigma_{\gamma\bar{t}c}$ and
$\sigma_{Z\bar{t}c}$ versus the mass $m_{\pi_{t}}$ for
$\sqrt{S}=500GeV$.
\item[Fig.6:]The cross sections $\sigma_{\gamma\bar{t}c}$ and
$\sigma_{Z\bar{t}c}$ versus $\sqrt{S}$ for $m_{\pi_{t}}=300GeV$.
\end{description}

\newpage
\begin{figure}[pt]
\begin{center}
\begin{picture}(250,200)(0,0)
\put(-50,0){\epsfxsize120mm\epsfbox{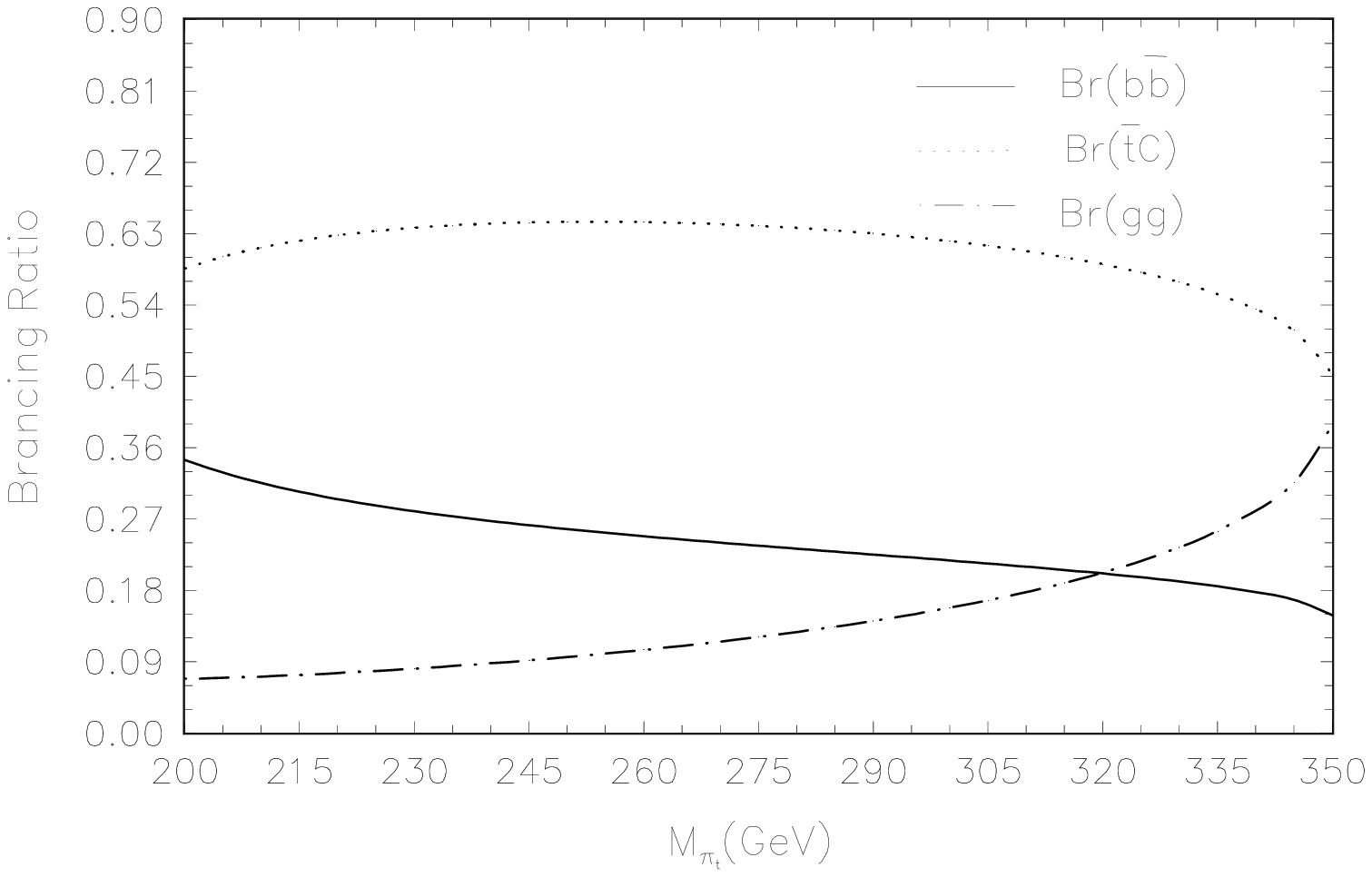}}
 \put(120,-10){Fig.1}
\end{picture}
\end{center}
\end{figure}

\begin{figure}[hb]
\begin{center}
\begin{picture}(250,200)(0,0)
\put(-50,0){\epsfxsize120mm\epsfbox{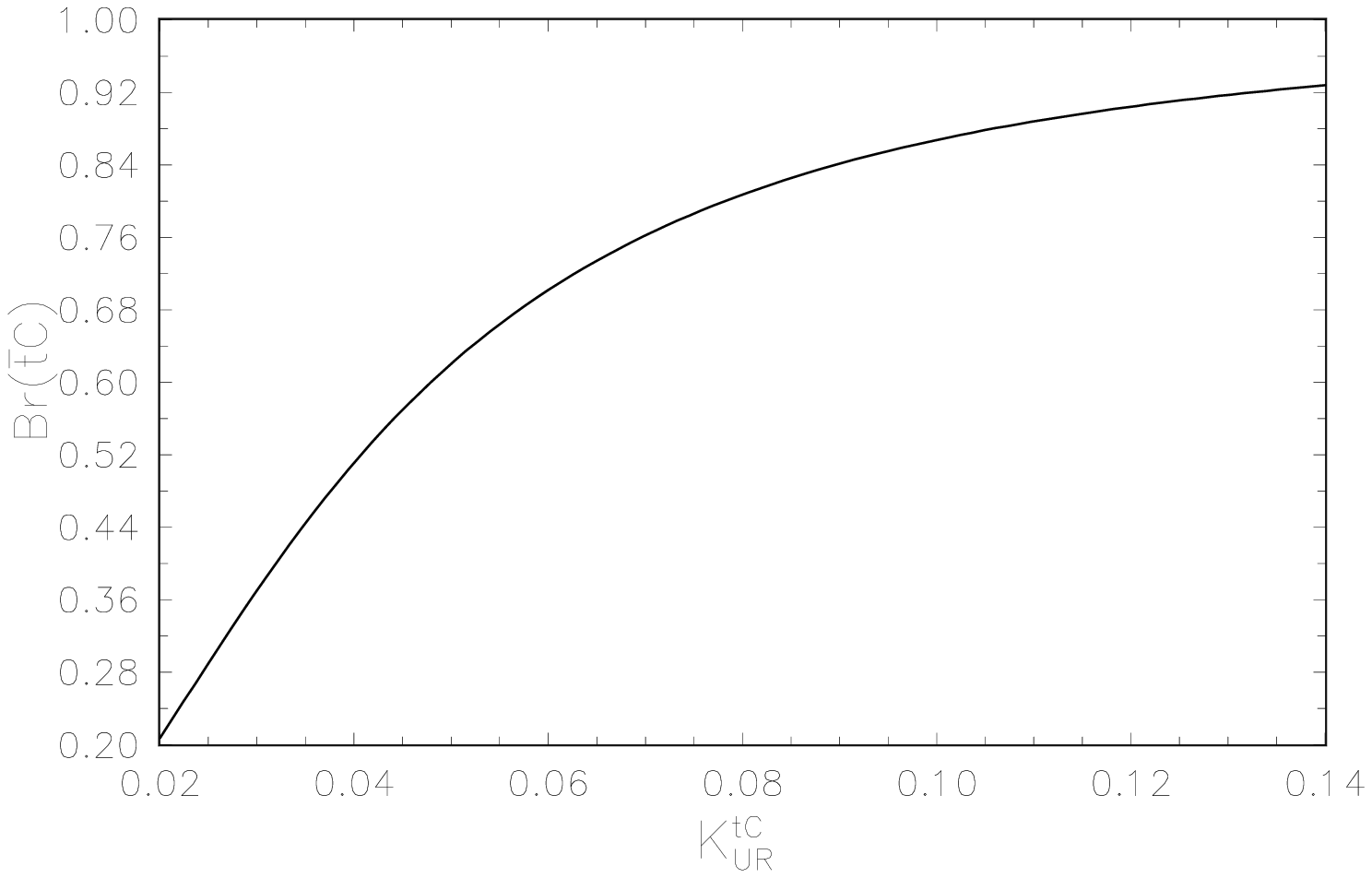}}\put(120,-10){Fig.2}
\end{picture}
\end{center}
\end{figure}

\newpage
\begin{figure}[pt]
\begin{center}
\begin{picture}(250,200)(0,0)
\put(-50,0){\epsfxsize120mm\epsfbox{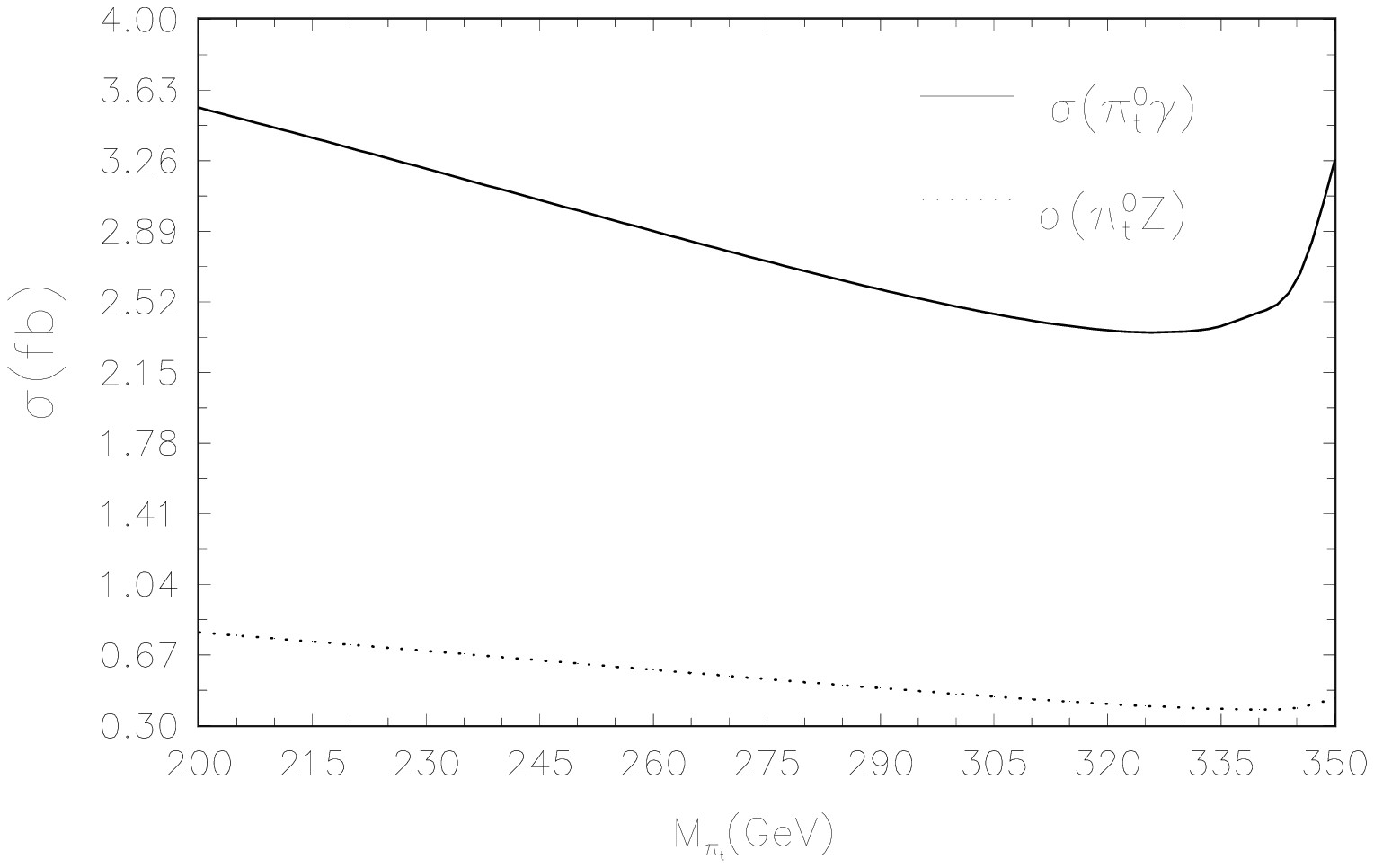}} \put(120,-10){Fig.3}
\end{picture}
\end{center}
\end{figure}

\begin{figure}[hb]
\begin{center}
\begin{picture}(250,200)(0,0)
\put(-50,0){\epsfxsize120mm\epsfbox{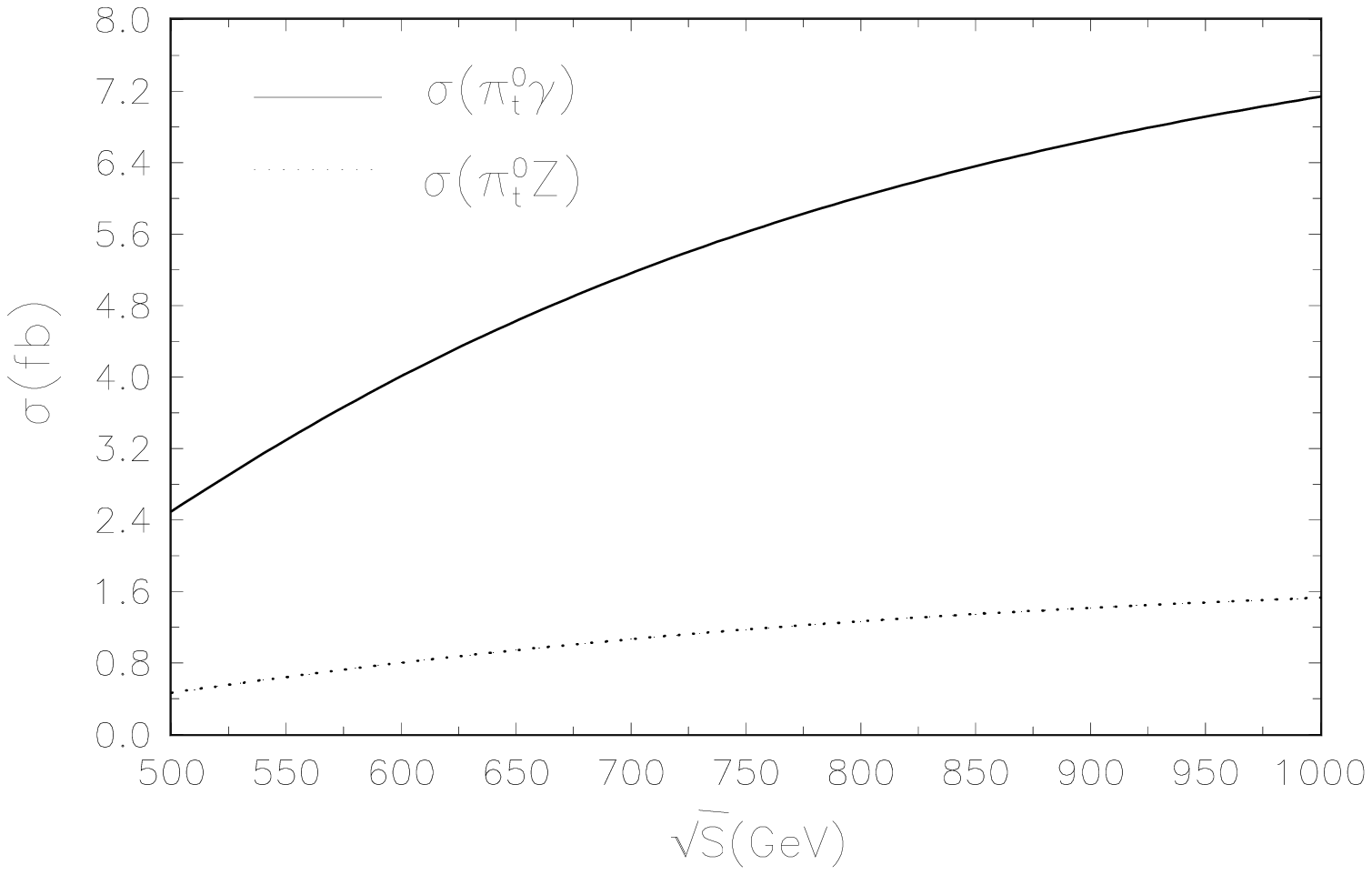}}\put(120,-10){Fig.4}
\end{picture}
\end{center}
\end{figure}

\newpage
\begin{figure}[pt]
\begin{center}
\begin{picture}(250,200)(0,0)
\put(-50,0){\epsfxsize120mm\epsfbox{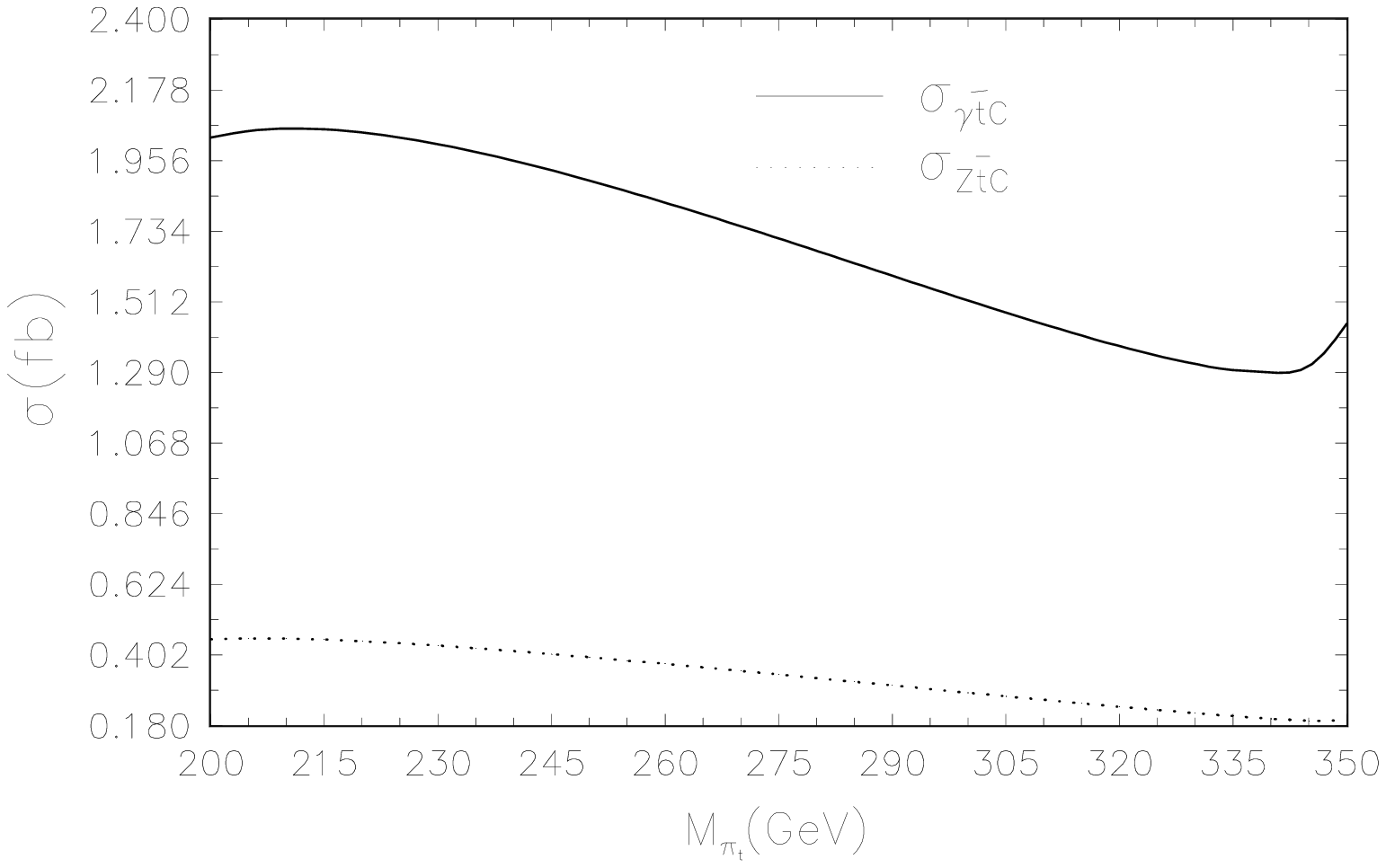}} \put(120,-10){Fig.5}
\end{picture}
\end{center}
\end{figure}

\begin{figure}[hb]
\begin{center}
\begin{picture}(250,200)(0,0)
\put(-50,0){\epsfxsize120mm\epsfbox{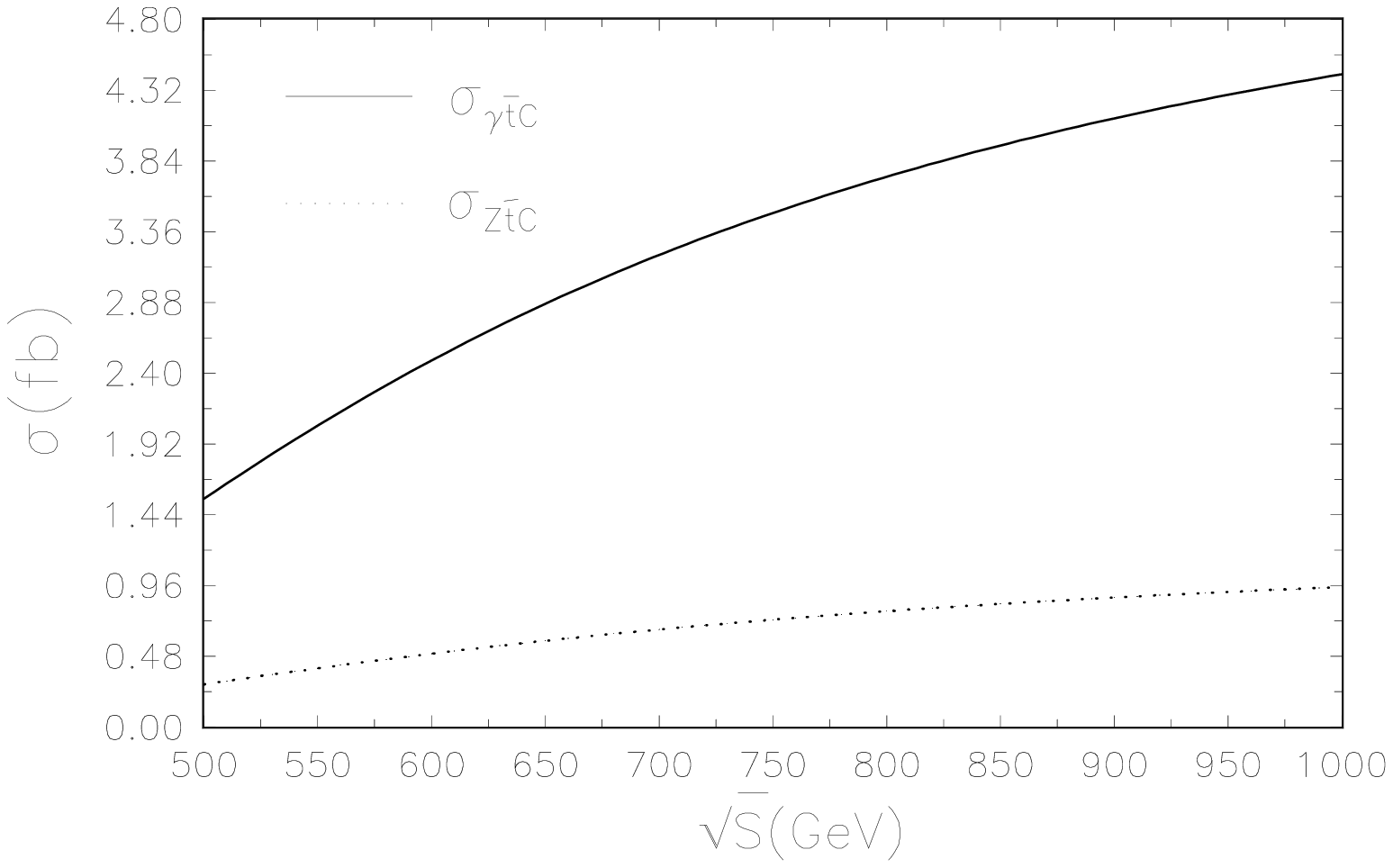}}\put(120,-10){Fig.6}
\end{picture}
\end{center}
\end{figure}

\newpage
\begin{thebibliography}{99}
\bibitem{x1} C.T.Hill, {\em Phys. Lett. B}{\bf 345}, 483(1995); K. Lane and
E. Eichten, {\em Phys. Lett. B}{\bf 352}, 382(1995); K. Lane, {\em
Phys. Lett. B}{\bf 433}, 96(1998).
\bibitem{x2} G.Buchalla, G. Burdman, C. T. Hill and D. Kominis,
{\em Phys. Rev. D}{\bf 53}, 5185(1996).
\bibitem{x3} V. Lubicz, {\em Nucl. Phys. B}{\bf 404}, 559(1993); V. Lubicz and
P. Santorelli, {\em Nucl. Phys. B}{\bf 460}, 3(1996); R.
Casalbuoni, A. Deandrea, S. D. Curtis, D. Dominici, R. Gatto, and
J. F. Gunion, {\em Nucl. Phys. B}{\bf 555}, 3(1999).
\bibitem{x4} R. S. Chivukula, R. Rosenfeld, E. H. Simmons and J.
Terning, hep-ph/9503202; K. Lane, {\em Phys. Lett. B}{\bf 357},
624(1995); E. Eichten, K. Lane and J. Womersley, {\em Phy. Lett.
B}{\bf 405}, 305(1997).
\bibitem{x5} E. Eichten and K. Lane, {\em Phys. Lett. B}{\bf 327}, 129(1994);
G. L. Lu, H. Yang, J. M. Yang, and X. L. Wang, {\em Phys. Rev.
D}{\bf 54}, 1083(1996); Chongxing Yue, Hongyi Zhou, Yuping Kuang,
and Gongru Lu, {\em Phys. Rev. D}{\bf 55}, 5541(1997).
\bibitem{x6} Hongyi Zhou, Yuping Kuang, Chongxing Yue and Hua Wang,
{\em Phys. Rev. D}{\bf 57}, 4205(1998).
\bibitem{x7} Xuelei Wang, et al, {\em Phys. Rev. D}{\bf 60}, 014002(1999).
\bibitem{x8} Hong-Jian He and C. P Yuan, {\em Phys. Rev. Lett}. {\bf 83},
28(1999); G. Burdman, {\em Phys. Rev. Lett}. {\bf 83}, 2888(1999).
\bibitem{x9} G. Burdman and D. Kominis, {\em Phys. Lett. B}{\bf 403},
101(1997); W. Loinaz and T. Takuchi, {\em Phys. Rev. D}{\bf 60},
015005(1999); Chongxing Yue, Yuping Kuang, Xuelei Wang, and Weibin
Li, {\em Phys. Rev. D}{\bf 62}, 055005(2000).
\bibitem{x10} Chongxing Yue, Yuping Kuang and Gongru Lu, {\em J. Phys.
G}{\bf 23}, 163(1997).
\bibitem{x11} D. Slaven, Bing-Ling Yang, and X. M. Zhang, {\em Phys.
Rev. D}{\bf 45}, 4349(1992); Chongxing Yue, Xuelei Wang, and
Gongru Lu, {\em J. Phys G}{\bf 19}, 821(1993).
\bibitem{x12} L. Randall and E. H. Simmons, {\em Nucl. Phys. B}{\bf 380},
3(1992).
\bibitem{x13} G. Eilam, {\em Phys. Rev. D}{\bf 18}, 1202(1983); C.H, Chang, et
al, {\em Phys. Lett. B}{\bf 313}, 389(1993); C. S. Huang, X. H.
Wu. and S. H. Zhu, {\em Phys. Lett. B}{\bf 452}, 143(1999).
\bibitem{x14} D. Atwood, L. Reina and A. Soni, {\em Phys. Rev.
D}{\bf 55}, 3156(1997); S. Bar-Shalom, G. Eilam, A. Soni and J.
Wudka, {\em Phys. Rev. Lett.}{\bf 79}, 1217(1997); ibid. {\em
Phys. Rev.D}{\bf 57}, 2957(1998); J. Yi et al, {\em Phys.
Rev.D}{\bf 57}, 4343(1998); W.-S. Hou, G.-L. Lin and C.-Y. Ma,
{\em Phys. Rev. D}{\bf 56}, 7434(1997); M. Sher, hep-ph/9809590.
\bibitem{x15} U. Mahanta and H. Ghosal, {\em  Phys. Rev. D}{\bf 57},
1735(1998); S. Bar-Shalom, G. Eilam and A. Soni, {\em Phys. Rev.
D}{\bf 59}, 035010(1999); M. Chemtob and G. Moreau, {\em Phys.
Rev. D}{\bf 59}, 116012(1999); Z.-H. Yu et al, {\em Eur. Phys. J.
C}{\bf 16}, 541(2000).
\bibitem{x16} F. del Aguila, J. A. Aguilar-Saavedra and R. Miquel,
{\em Phys. Rev. Lett}. {\bf 82},1628(1999).
\bibitem{x17} T. Han and J. L. Hewett, {\em Phys. Rev. D}{\bf 60},
074015(1999); S. Bar-Shalom and J. Wudka, {\em Phys. Rev. D}{\bf
60}, 094016(1999).
\bibitem{x18} Chongxing Yue and Weibin Li, {\em Mod. Phys.
Lett.A}{\bf 15}, 361(2000); Chongxing Yue, et al,
hep-ph/0011112(to be published in {\em Phys. Lett. B}).
\end{thebibliography}
\end{document}